\documentclass[journal]{IEEEtran}
\usepackage{url}
\usepackage[utf8]{inputenc}
\usepackage{xcolor}
\usepackage{amsmath}
\usepackage{amssymb}

\usepackage{tablefootnote}
\usepackage{booktabs}
\usepackage{tabularx}
\usepackage{tikz}
\usepackage{pgfplots}
\pgfplotsset{compat=newest} 
\pgfplotsset{plot coordinates/math parser=false} 
\newlength\fheight
\newlength\fwidth
\usetikzlibrary{plotmarks,patterns,decorations.pathreplacing,backgrounds,calc,arrows,arrows.meta,spy,matrix}
\usepgfplotslibrary{patchplots,groupplots}
\usepackage{tikzscale}
\usepackage{siunitx}

\usepackage{tikz}
    \usetikzlibrary{shapes.arrows}

\usepackage{pgfplots}
\usepackage{adjustbox}

\usepackage{gincltex}
\usepgfplotslibrary{fillbetween}
\usetikzlibrary{plotmarks}

\usepackage{multirow}
\usepackage[font=scriptsize]{subcaption}
\usepackage[font=footnotesize]{caption}

\usepackage{mathtools}

\usepackage{dblfloatfix}    
\usepackage{colortbl}

\usepackage{makecell}
\usepackage{diagbox}
\usepackage{tikz-qtree}
\usetikzlibrary{trees} 
\usepackage{cite}
\usepackage{bbold}
\usepackage{bbm}

\usepackage{array}
\newcolumntype{?}{!{\vrule width 1.5pt}}
\newcolumntype{P}[1]{>{\centering\arraybackslash}p{#1}}
\usepackage{makecell}
\usepackage{mdframed}
\usepackage[many]{tcolorbox}
\usepackage{enumitem}
\usepackage{booktabs}

\definecolor{violet}{rgb}{0.6,0,0.6}%
\definecolor{orange_D}{rgb}{1,0.3,0}%
\definecolor{green_D}{rgb}{0,0.6,0}%
\definecolor{cyan}{rgb}{0,0.67,0.64}%
\definecolor{red}{rgb}{0.9,0,0}%
\definecolor{green}{rgb}{0,0.8,0}%
\definecolor{yellow}{rgb}{1,0.8,0}
\definecolor{color0}{HTML}{FFD700}
\definecolor{color1}{HTML}{FFB14E}
\definecolor{color2}{HTML}{FA8775}
\definecolor{color3}{HTML}{EA5F94}
\definecolor{color4}{HTML}{CD34B5}
\definecolor{color5}{HTML}{9D02D7}
\definecolor{color6}{HTML}{0000FF}

\newtcbox{\mybox}[1][]{nobeforeafter,math upper,tcbox raise base,
  enhanced,frame hidden,boxrule=0pt,interior style={top color=green!10!white,
  bottom color=green!10!white,middle color=green!50!yellow},
  fuzzy halo=1pt with green,drop large lifted shadow,#1}

\def \fwidth{0.8\columnwidth}
\def \fheight {0.4\columnwidth}

\usetikzlibrary{fadings}

\tikzfading[name=middle,
            top color=transparent!100,
            bottom color=transparent!100,
            middle color=transparent!20]

\usetikzlibrary{arrows,automata,calc,shapes, positioning,shadows,shadows.blur,shapes.geometric}

\usepackage[acronyms,nonumberlist,nopostdot,nomain,nogroupskip]{glossaries}
\newacronym{cbr}{CBR}{Constant Bit Rate}
\newacronym[\glslongpluralkey={Markov Decision Processes}]{mdp}{MDP}{Markov Decision Process}
\newacronym{pdf}{PDF}{Probability Density Function}
\newacronym{cdf}{CDF}{Cumulative Distribution Function}
\newacronym{mc}{MC}{Markov Chain}
\newacronym{mcs}{MCS}{Modulation and Coding Scheme}
\newacronym{edf}{EDF}{Earliest Deadline First}
\newacronym{hlf}{HLF}{Highest Level First}
\newacronym{qos}{QoS}{Quality of Service}
\newacronym{mptcp}{MPTCP}{Multi-path TCP}
\newacronym{lowrtt}{LowRTT}{lowest RTT first}
\newacronym{dems}{DEMS}{Decoupled Multipath Scheduler}
\newacronym{stms}{STMS}{Slide Together Multipath Scheduler}
\newacronym{daps}{DAPS}{Delay Aware Packet Scheduling}
\newacronym{blest}{BLEST}{Blocking Estimation}
\newacronym{iid}{IID}{Independent and Identically Distributed}
\newacronym{rtt}{RTT}{Round-Trip Time}
\newacronym{fec}{FEC}{Forward Error Correction}
\newacronym{leap}{LEAP}{Latency-controlled End-to-End Aggregation Protocol}
\newacronym{urllc}{URLLC}{Ultra-Reliable Low Latency Communications}
\newacronym{ccr}{CCR}{Constant Coding Rate}
\newacronym{ps}{PS}{Plain Split}
\newacronym{pec}{PEC}{Packet Erasure Channel}
\newacronym{pmf}{PMF}{Probability Mass Function}
\newacronym{hop}{HOP}{High-reliability latency-bounded Overlay Protocol}
\newacronym{vr}{VR}{Virtual Reality}
\newacronym{ar}{AR}{Augmented Reality}
\newacronym{qs}{QS}{Queuing System}
\newacronym{iot}{IoT}{Internet of Things}
\newacronym{v2x}{V2X}{Vehicle to Everything}
\newacronym{ran}{RAN}{Radio Access Network}
\newacronym{rssi}{RSSI}{Received Signal Strength Indicator}
\newacronym{bs}{BS}{Base Station}
\newacronym{arq}{ARQ}{Automatic Repeat reQuest}
\newacronym{harq}{HARQ}{Hybrid Automatic Repeat reQuest}
\newacronym{hol}{HoL}{Head of Line blocking}
\newacronym{bbr}{BBR}{Bottleneck Bandwidth and Round-trip propagation time}
\newacronym{fps}{FPS}{Frames per Second}
\newacronym{snr}{SNR}{Signal to Noise Ratio}
\newacronym{rl}{RL}{Reinforcement Learning}
\newacronym{pomdp}{POMDP}{Partially Observable MDP}

\usepackage{hyperref}

\begin{document}
\pagenumbering{gobble}

\title{Joint Scheduling and Coding for Reliable, Latency-bounded Transmission over Parallel Wireless Links}


\author{{Andrea Bedin,~\IEEEmembership{Student Member, IEEE}, Federico Chiariotti,~\IEEEmembership{Member, IEEE}, Andrea Zanella,~\IEEEmembership{Senior Member, IEEE}}
\thanks{
Andrea Bedin (andrea.bedin@nokia.com) and Andrea Zanella (zanella@dei.unipd.it) are with the Department of Information Engineering, University of Padova, Italy. Andrea Bedin is also with Nokia Bell Labs, Espoo, Finland.
Federico Chiariotti (corresponding author, fchi@es.aau.dk) is with the Department of Electronic Systems, Aalborg University, Denmark. This project has received funding from the European Union's Horizon 2020 research and innovation program under the Marie Skłodowska-Curie Grant agreement No. 861222.
}
}

\maketitle

\begin{abstract}
Several novel industrial applications involve human control of vehicles, cranes, or mobile robots through various high-throughput feedback systems, such as \gls{vr} and tactile/haptic signals. The near real-time interaction between the system and the operator requires strict latency  constraints in packet exchange, which is difficult to guarantee over wireless communication links. In this work, we advocate that packet-level coding and packet scheduling over multiple parallel (unreliable) links have the potential to provide reliable, latency-bounded communication for applications with periodic data generation patterns. However, this goal can be reached only through a careful joint design of such mechanisms, whose interactions can be subtle and difficult to predict. In this paper we first discuss these aspects in general terms, and then present a \gls{mdp} model that can be used to find a scheme that optimally exploits the multichannel wireless access in order to maximize the fraction of data blocks delivered within deadline. Our illustrative example is then used to show the optimal coding/scheduling strategies under different combinations of wireless links, also showing that the common solution of backing up a high bitrate unreliable mmWave link with a low bitrate more stable sub-6~GHz link can actually be ineffective in the considered scenario. 
\end{abstract}


\glsresetall

\section{Introduction}
\label{sec:introduction}

The new Industry~4.0 paradigm has led to the automation and digitization of many industrial processes, with an increased emphasis on wireless communication technologies: in several industrial and commercial scenarios, the greater flexibility and mobility offered by wireless connections is crucial to perform tasks effectively. The evolution of mobile networks, first with 5G and now with the first steps towards 6G, caters specifically to industrial use cases through the definition of the \gls{urllc} traffic class~\cite{saad2019vision}: the extremely high reliability and predictable latency of cellular wireless access can enable the safe operation of machinery without having to consider communication system faults and imperfections in the control design. However, the stringent constraints of this traffic class require a significant amount of resources, as well as a high level of support from the network: without scheduling resources in advance and planning transmissions to avoid interference, delivering packets with extremely low latency becomes an arduous task.

In addition, some applications involve  mobility and high throughput, making resource allocation even more challenging. A significant example is the remote maneuvering of mobile devices, such as robotic arms, cranes, forklifts, or pickup trucks inside a warehouse, or at the docks, in a mixed environment with human workers~\cite{TS22104}. While basic safety mechanisms have to be implemented locally to avoid accidents (e.g., emergency breaking in case of nearby obstacle), macro maneuvering can be left to a remote (human) operator, possibly equipped with a \gls{vr} visor or, more simply, a screen to visualize the scene around the robot and take appropriate actions, which are then passed to the robot~\cite{hu2021vision}. 

In this scenario, high-quality video needs to be streamed to the control station with minimal delay to allow for real-time control~\cite{gholipoor2020cloud}. For example, the IEEE~\cite{ieee2021vrstandard} specifications for high-performance teleoperation require a 20~ms round-trip delay and 4K video resolution, as better explained in Sec.~\ref{sec:sim}. However, the variability of the wireless channel due to nodes' mobility makes static resource allocation unsuitable to guarantee the strict latency requirements of the application, if not by over-provisioning the communication channel, which is neither efficient nor scalable.  At the same time, the consequences of late delivery of data may be significant, since the lack of control information may trigger safety mechanisms that stop the robot (also abruptly) to avoid accidents, yielding waste of time and, possibly, mechanical stresses in case of emergency breaking of heavy equipment. Therefore, an unreliable connection can make the whole application unworkable.

The aggregation of multiple wireless links is an interesting option to provide reliability by putting together unreliable components: failures on one link can be compensated by the others, making the connection as a whole stronger and more reliable. This principle has been exploited at the physical and medium access layers, aggregating the transmissions from multiple \glspl{bs} or over multiple subcarriers to strengthen the received signal and exploit the independence of wireless channels~\cite{nielsen2017ultra}. However, this has significant signaling and coordination requirements, and can only be performed with full control of the network. It is also possible to aggregate multiple wireless channels \emph{end-to-end}, considering the network as a black box~\cite{chiariotti2021hop}. This approach is significantly more flexible, as it can be applied over any network or combination of networks, aggregating even different operators and technologies, but still needs careful optimization, as there are several potential issues.

\begin{figure*}[t]
     \centering
     \input{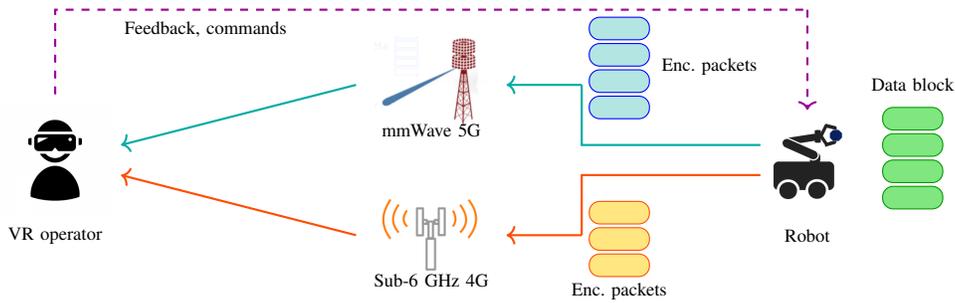}
     \caption{Schematic of VR teleoperation over a mixed 4G/5G multipath connection.}
     \label{fig:schematic}
\end{figure*}

In this work, we present the main issues and solutions for multipath aggregation, aimed at real-time control use cases with high throughput, significant reliability requirements, and strict application-level latency constraints. The challenges presented by these use-cases cannot be solved by purely access network mechanisms, such as \gls{urllc}, as performance has to be evaluated end-to-end, not only on the wireless segment. 
The solution we propose, then, focuses on application data blocks (e.g., video frames), which can be protected using packet-level network coding and transmitted over multiple independent connections. The transmission of each block can be optimized as a \gls{mdp}, a model which can capture the effect of sequential actions on the future state of the network~\cite{howard1960dynamic}.
We discuss the main issues affecting this type of strategies, as well as the existing support in currently standardized protocols and networks, and present the main trade-offs in a remote-control use case.

The rest of this paper is organized as follows: first, we discuss the requirements and challenges of reliable low-latency multipath transmission in Sec.~\ref{sec:requirements}. We then present a possible solution to the scheduling and packet-level coding optimization problem in Sec.~\ref{sec:mdp}, which we evaluate in  a mixed mmWave and sub-6~GHz scenario in Sec.~\ref{sec:sim}. Finally, we conclude the paper and present some potential research directions in Sec.~\ref{sec:conc}.

\section{Reliable Multipath Transmission}\label{sec:requirements}

We consider a throughput-intensive application generating data blocks at regular intervals. Remote control applications involving rich feedback, such as video or \gls{ar}, generate this type of traffic: frames are recorded or rendered at a constant pace, but might have different sizes, generally much larger than other sensor data in industrial applications.

We then have two potential scenarios, which depend on the nature of the wireless channels being used:
\begin{itemize}
    \item \emph{Time-varying} channels have a variable capacity, which can fluctuate depending on the nodes' mobility and propagation characteristics of the environment, as well as on the cross-traffic sharing the same wireless resources. Depending on the amount of data transmitted through the wireless link, channel fluctuations might result in higher delivery latency or packet loss rates;
    \item \emph{On/off} channels can either deliver all the data transmitted through them within the deadline, or be completely unavailable. Links at mmWave frequencies are likely the most significant example of this behavior in modern networks~\cite{zhang2019will}, as blockages can make the direct link completely unavailable.
\end{itemize}
In both cases, a single wireless link cannot guarantee reliable transmission within a tight latency constraint: time-varying channels might only deliver part of the data on time, while on-off channels might drop the entire frame, but the overall effect on the application is the same. Traditional end-to-end protocols such as TCP or QUIC deal with this issue using \gls{arq}, i.e., waiting for the packets to be acknowledged and retransmitting lost or excessively delayed~\cite{polese2019survey} data. However, retransmissions have a significant impact on the frame delay, and can cause the \gls{hol} problem: since packets and blocks need to be delivered in the correct order, packet loss can negatively affect even future frames, as it will prevent the application from getting the data until the retransmission is successful.

In these cases, packet-level \gls{fec} is a powerful alternative, exploiting redundant information to allow the receiver to reconstruct missing data. On a single channel, this can mitigate the effects of packet loss, but it is ineffective if the issue is a drop in the link's capacity, since the block transmission with \gls{fec} will take even longer, increasing the delay and the risk of violating the latency constraint. Conversely, \gls{fec} can help when using multiple parallel links~\cite{chiariotti2021hop}: if the channels are independent, blockages or capacity drops on one link can be compensated by the others, if the redundancy is sufficient. 

Fig.~\ref{fig:schematic} shows an example of the effect of multipath coding in the context of a teleoperation task: an industrial robot observes the environment and renders its camera data into a \gls{vr} feed, which needs to be transmitted to the user. The data block in the example is composed of 4 original (green) packets that are encoded as 7 coded packets: four packets (cyan) are transmitted on a mmWave 5G link and three (orange) on a backup 4G sub-6 link for redundancy and reliability. The reception of any set of 4 packets out of the 7 transmitted will allow the receiver to recover the original frame and display it to the human operator, increasing the probability of meeting the deadline.

However, exploiting path diversity to ensure reliability poses a difficult challenge: since transmitting more packets on a link can increase queuing and, hence, delay for future frames, the amount of redundancy and the splitting of coded packets among the available links need to be optimized to guarantee both high reliability for the current frame and a relatively low impact on future frames.

\subsection{Self-Queuing Delay} \label{sec:sqdelay}
Queuing delay is a key issue for time-constrained applications: if the capacity of a link is exceeded, packets pile up, increasing the delay not only for the current frame, but also for future ones. The self-queuing delay issue is well-known in the end-to-end transport literature~\cite{polese2019survey}, and led to the design of several low-latency TCP versions.

We can distinguish three cases, which depend on the nature of the application and on the amount of available network support:
\begin{enumerate}
    \item In \emph{underloaded} links, the capacity of the link is much higher than the application bitrate, and queuing is extremely rare. In this case, self-queuing is not an issue, and even robust coding can be supported easily. When available (i.e., not blocked), the mmWave links can be considered as underloaded;
    \item In \emph{preemptive} links, the link is under high load, but the application can control whether packets in the queue are transmitted or discarded. If the previous block has been received correctly, or can be superseded by the next one (which contains more up-to-date information), queuing is not an issue, as older packets are simply dropped from the queue. However, if the older block is still being transmitted, and is required for decoding the new information (as in most video encoding schemes), queuing can become a problem. Furthermore, preemptive operation requires a high level of integration with the network, which might not always be available;
    \item In \emph{uncontrolled} links, packets that are sent through the link are only discarded in case of buffer overflows, and congestion can have a significant impact on the delay of future frames. This is the case in most public connections, in which the networking infrastructure is not directly controlled by the system designer.
\end{enumerate}
These three cases are increasingly complex to optimize, as actions have deeper consequences on the future viability of the link: while transmitting too many packets through an underloaded link has almost no consequences, doing the same over an uncontrolled link can affect multiple future frames, causing a significant performance drop~\cite{shah2016redundant}.

\subsection{Link Quality Estimation}\label{ssec:estimate}

In order to effectively determine the amount of redundancy needed to transmit a frame reliably and within the deadline, we need to have an estimate of the distribution of future link capacity: in on/off links, this is equivalent to estimating the blockage probability, as packets are always delivered on time when the link is available. On the other hand, loaded, time-varying links present a tougher challenge: in order to optimize the redundancy, we need an estimate of the \gls{pmf} of the number of packets that can be delivered over them in a certain time interval. This might be partly obtained by querying the network interface firmware for the link quality information, but cross-traffic is harder to estimate.

In general, a higher uncertainty on this \gls{pmf} is particularly damaging for uncontrolled links, as it increases the amount of redundancy necessary for full reliability, inevitably affecting future frames due to self-queuing delay. In addition, network support is less likely for uncontrolled links, so the link quality estimation might have to be performed entirely over the top, with no direct information on the signal quality or state of the buffer.

Over the top estimation mechanisms use ACK packets to infer the available end-to-end connection capacity without any cross-layer input, and are exploited by congestion control algorithms such as Google's \gls{bbr}~\cite{atxutegi2018use}. However, they also present some additional challenges, as the scheduling process and the capacity estimation process are inherently intertwined:
\begin{itemize}
    \item Delayed feedback might have a significant effect on the estimation of both the state of the queue and the channel capacity: as we are considering individual wireless links, this delay might not be large, but it needs to be taken into account;
    \item Estimating large capacities requires bursty transmission to saturate the channel, which may lead to higher delays. \gls{bbr} periodically increases its pace to detect capacity increases, but is relatively slow to adapt to stepwise changes, which can be relatively common in  wireless links, e.g., due to changes in the \gls{mcs}; 
    \item Low-capacity or high-uncertainty links are typically under-used by the scheduler, as they do not provide any guarantees of timely delivery. However, the quality of the estimate also depends on a minimum number of packets being sent to get accurate statistics. As sending fewer packets increases uncertainty, and increased uncertainty leads schedulers to reduce the number of packets sent through the link, the feedback loop can snowball into making high-uncertainty links entirely unusable.
\end{itemize}

\section{Joint Coding and Scheduling Optimization}\label{sec:mdp}

We then consider a solution to jointly optimize the coding and scheduling of the blocks.
Scheduling more packets on a given link has two effects: the first is that the probability of delivering the block on time increases thanks to the additional redundancy, and the second is that the probability of generating a queue for the next block increases (if the link is uncontrolled). These may not even be at the same time, as the deadline for a block is not necessarily tied to the generation of the next block: \gls{vr} frames are usually transmitted at very high frame rates, accommodating longer delays in the network, so that multiple frames might be in the transmission buffer at the same time. However, the only moment when the transmitter needs to make a decision is when a new frame is generated, and most applications follow deterministic or at least partially predictable patterns.

\begin{figure}[t]
     \centering
     \begin{tikzpicture}

\node[draw,minimum height=1cm,minimum width=3cm] (env) at (0,0) {\scriptsize Environment (channels)};

\node[draw,minimum height=1cm,minimum width=3cm] (agent) at (0,3) {\scriptsize Scheduler};

 \draw[->] (-3,0.1) |- ([yshift=-0.1cm]agent.west);
 \draw[-] (-3,0.1) -- ([yshift=0.1cm]env.west);
  \draw[->] (-3.2,-0.1) |- ([yshift=0.1cm]agent.west);
 \draw[-] (-3.2,-0.1) -- ([yshift=-0.1cm]env.west);
 \node[anchor=west] (fb1) at (-3,1.7) {\scriptsize Link quality, queue state};
 \node[anchor=west] (fb2) at (-3,1.3) {\scriptsize  \emph{(state)}};
 \node[anchor=east] (rw1) at (-3.2,1.7) {\scriptsize  Block ACK};
 \node[anchor=east] (rw2) at (-3.2,1.3) {\scriptsize  \emph{(reward)}};
 \node[anchor=east] (act) at (3,1.5) {\scriptsize  Packet schedule \emph{(action)}};
 
\draw[->] (3,3) |- (env.east);
 \draw[-] (3,3) -- (agent.east);

\node[draw,rounded corners,minimum height=1cm,minimum width=2.25cm] (s00) at (-3,-3) {};

 \node (s001) at (-3,-2.8) {\scriptsize  State: (good, good)};
 \node (s002) at (-3,-3.2) {\scriptsize  Action: $(1,1)$};

\node[draw,rounded corners,minimum height=1cm,minimum width=2.25cm] (s10) at (0,-1.75) {};

 \node (s101) at (0,-1.55) {\scriptsize  State: (good, bad)};
 \node (s102) at (0,-1.95) {\scriptsize  Action: $(1,0)$};

\node[draw,rounded corners,minimum height=1cm,minimum width=2.25cm] (s01) at (0,-4.25) {};

 \node (s011) at (0,-4.05) {\scriptsize  State: (bad, good)};
 \node (s012) at (0,-4.45) {\scriptsize  Action: $(0,1)$};

\node[draw,rounded corners,minimum height=1cm,minimum width=2cm] (s11) at (3,-3) {};

 \node (s011) at (3,-2.8) {\scriptsize  State: (bad, bad)};
 \node (s012) at (3,-3.2) {\scriptsize  Action: $(0,0)$};
 
\draw[->] ([xshift=0.1cm]s00.south) |- node [above,near end] {\scriptsize$p=0.08$} ([yshift=0.1cm]s01.west);
\draw[<-] ([xshift=-0.1cm]s00.south) |- ([yshift=-0.1cm]s01.west);

\draw[<-] ([xshift=-0.1cm]s00.north) |-  ([yshift=0.1cm]s10.west);
\draw[->] ([xshift=0.1cm]s00.north) |- node [below,near end] {\scriptsize$p=0.18$} ([yshift=-0.1cm]s10.west);

\draw[->] ([yshift=0.1cm]s00.east) -- node [above,near start] {\scriptsize$p=0.02$} ([yshift=0.1cm]s11.west);
\draw[<-] ([yshift=-0.1cm]s00.east) -- ([yshift=-0.1cm]s11.west);

\draw[->] ([xshift=-0.1cm]s11.south) |- ([yshift=0.1cm]s01.east);
\draw[<-] ([xshift=0.1cm]s11.south) |-  ([yshift=-0.1cm]s01.east);

\draw[->] ([xshift=-0.1cm]s10.south) -- ([xshift=-0.1cm]s01.north);
\draw[<-] ([xshift=0.1cm]s10.south) --  ([xshift=0.1cm]s01.north);

\draw[<-] ([xshift=0.1cm]s11.north) |- ([yshift=0.1cm]s10.east);
\draw[->] ([xshift=-0.1cm]s11.north) |- ([yshift=-0.1cm]s10.east);

\draw[-] (-4.125,-2.75) to [out=180,in=270] (-4.375,-2.5);
\draw[-] (-4.375,-2.5) to [out=90,in=180] (-4.125,-2.25);
\draw[->] (-4.125,-2.25) to [out=0,in=90] (-3.875,-2.5);

 \node (st) at (-4,-2.05) {\scriptsize $p=0.72$};

\end{tikzpicture}
     \caption{Above: Schematic of VR tele-operation over a mixed 4G/5G multipath connection. Below: example of system model, with two good/bad (Gilbert-Elliot) wireless channels and single-packet blocks.}
     \label{fig:mdp}
\end{figure}
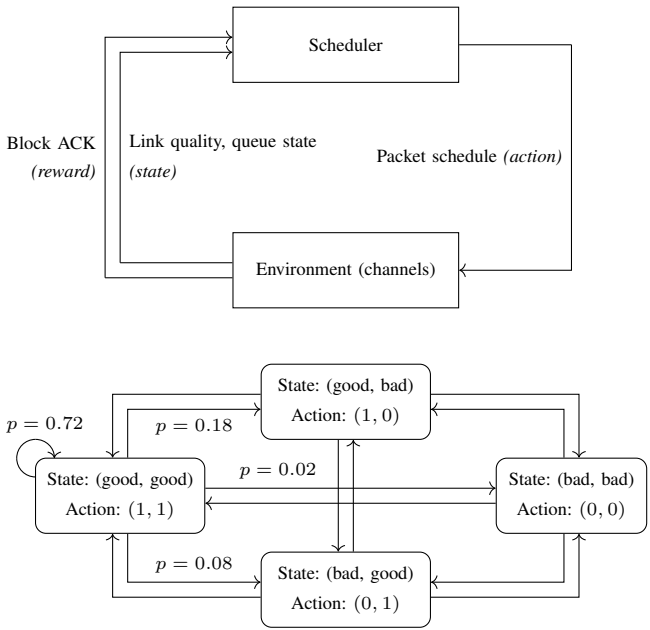

\glspl{mdp} are a natural way to model systems making sequential decisions whose consequences can affect future performance in potentially complex ways. The underlying model is a Markov chain, i.e., a stochastic process that takes values in a \emph{discrete state} space and whose future evolution does not depend on the past history. The transition from one state to the next is stochastic and depends on transition probabilities that are affected by the \emph{strategy} adopted by an \emph{agent}, who can observe the current state and take \emph{actions} in a certain set, receiving a \emph{reward} of some kind. The final objective is to find the strategy that maximizes the long-term expected reward. The literature on \glspl{mdp} is extensive, and there are several methods to solve them and find the optimal \emph{policy} mapping each state to the best possible action in that state: usually, only small \glspl{mdp} in which the system model is well-known \emph{a priori} can be optimized analytically, and most of the literature focuses on \gls{rl} solutions that learn a policy from experience by trial and error~\cite{xiang2021recent}.

A natural representation of the reliable multipath communication problem as an \gls{mdp} is shown in the upper part of Fig.~\ref{fig:mdp}, with the following main components:
\begin{itemize}
    \item The \emph{actions} of the system correspond to the possible packet schedules among the different links: in the graphical example in Fig.~\ref{fig:schematic}, encoding the 4 original packets into 7, sending 4 packets over the mmWave link and 3 on the 4G sub-6 one, is a possible action. 
    If the connection includes preemptive links, dropping packets from the queue also needs to be considered as a potential component of the agent's action set. Similarly, if all links are uncontrolled, dropping a block outright is a potential action that can help ease congestion, at the cost of sacrificing a single block;
    \item The \emph{reward} function is the on-time delivery of the (decoded) block, often communicated through a block-level acknowledgment, regardless of the links the packets were delivered through. As mentioned, most \gls{mdp} solutions try to maximize the expected reward over long timescales, ensuring that the agent will not act myopically;
    \item The \emph{state} of the system is the most complex element, as it includes information about the physical state of each wireless channel, the cross-traffic on the link, and the state of the transmission buffer. The state of each wireless channel can possibly be summarized in a single value, such as the \gls{snr} or the \gls{rssi} or similar, but the uncertainty will increase with the inaccuracy of the information, leading to a lower reliance on the uncertain link. In on/off links, this is simplified to a single value representing the blockage probability.
\end{itemize}

The transition probability from one state to the other is naturally very complex, and may be impossible to model in realistic systems, leading to the use of \gls{rl} to find the optimal solution. The lower part of Fig.~\ref{fig:mdp} shows the \gls{mdp} model of a simple scenario with single-packet data blocks and two parallel binary (good/bad) wireless channels. The system state is hence given by the state of both channels and the action indicates the number of packets to be sent on each channel. 

However, the basic model is the same for a complex network with several links of different types, as the only difference is the size of the state and action spaces of the problem. Additionally, we only consider cases in which the state of the link can be estimated directly from the network interface: the case in which estimation is entirely performed end-to-end, as described in Sec.~\ref{ssec:estimate}, adds a third effect to the agent's action, which is to provide an estimate of the state. Systems in which the state is not observed directly and ideally, but its accuracy depends on the agent's actions, are modeled as \glspl{pomdp}, a more complex extension of \glspl{mdp}. Strategies also become slightly different: for example, it might be convenient to transmit a few packets over a link known to be in outage, as it is the only way to infer when the outage condition is over and the link becomes usable again~\cite{xiang2021recent}.

\section{Model Settings and Results}\label{sec:sim}

We consider a scenario in which a robot is remotely controlled by a \gls{vr} operator. The maximum motion-to-photon latency to avoid cybersickness and loss of performance is defined in a recent IEEE standard~\cite{ieee2021vrstandard} as 20~ms, which also specifies a required frame rate of at least 120~\gls{fps}. This leaves about 12.5~ms for the transmission of the \gls{vr} feed, considering realistic delays in the frame generation and feedback transmission. The standard also recommends 4K resolution, which results in a bitrate of approximately 48~Mb/s using the advanced H.265 compression standard~\cite{akyazi2018comparison}. Each frame is then approximately 400~kb, and we can consider encoder settings that result in approximately \gls{cbr} streaming. The main metric we aim at optimizing is the latency-constrained reliability, i.e., the probability that a new \gls{vr} frame will be delivered within the deadline.

We compare three different cases with two parallel wireless channels, which are assumed to be statistically independent:
\begin{itemize}
    \item \emph{Pure mmWave}: the robot can transmit the video feed to two \glspl{bs} in parallel. The channels are considered as on/off, as mmWave has a very high capacity when in line-of-sight, but has frequent outages due to blockages. The channels are then defined by their outage probabilities ($p_{\rm out}$), since the mean sojourn times in such a state are fixed to 5 steps;
    \item \emph{Pure sub-6}: the robot can transmit the video over two independent channels in the traditional frequency bands for cellular networks, commonly used for 4G and 5G. These channels are less affected by blockages and other physical obstacles, as lower frequencies have a better penetration, but can easily become congested, as the lower bandwidth can not handle all the cross-traffic. As commonly seen in the literature, the transmission over a sub-6 channel is modeled as an exponentially distributed time, whose parameter is given by the mean link capacity $C_{\rm Sub6}$ divided by the packet size;
    \item \emph{Mixed}: this is a commonly proposed solution to the problem of mmWave blockage, with a sub-6 channel serving as a backup when there is no line-of-sight mmWave link ~\cite{liu2018mec}.
\end{itemize}
The wireless channel models are abstract, as they depend on a single parameter (the outage probability for mmWave links and the average capacity for sub-6 channels). In addition, we assume the transmitter instantaneously knows the conditions of the wireless channels (assuming, e.g., cross-layer communication with the wireless cards or accurate channel-state estimation mechanisms). Such assumptions allowed us to derive the optimal policy and its performance analytically. In a more realistic scenario, this would possibly require more complex solutions such as \gls{rl}.

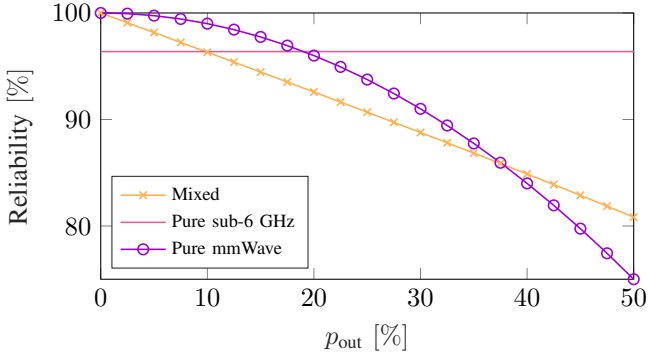
\begin{figure}[t]
  \centering
%
%

%
\begin{tikzpicture}

\begin{axis}[%
width=\fwidth,
height=\fheight,
scale only axis,
xmin=0,
xmax=50,
xlabel style={font=\color{white!15!black}},
xlabel={$p_{\text{out}}\ [\%]$},
ymin=75,
ymax=100,
ylabel style={font=\color{white!15!black}},
ylabel={Reliability $[\%]$},
axis background/.style={fill=white},
legend style={legend cell align=left, align=left, draw=white!15!black, at={(0.02,0.04)},anchor=south west,font={\scriptsize}}
]

\addplot [color=color1, semithick, mark=x, mark options={solid, color1}]
  table[row sep=crcr]{%
 0 100\\
2.5	99.0824574074383\\
5	98.1625610287537\\
7.5	97.2401037103913\\
10	96.3148532976954\\
12.5	95.3865487525956\\
15	94.4548955260177\\
17.5	93.5195600128636\\
20	92.5801628706071\\
22.5	91.6362709207071\\
25	90.687387270169\\
27.5	89.7329391806521\\
30	88.7762947170638\\
32.5	87.8137817378625\\
35	86.8447071234562\\
37.5	85.8681112958344\\
40	84.8829543621122\\
42.5	83.8879965681\\
45	82.8817481436135\\
47.5	81.8624033391255\\
50	80.8277525557979\\
};
\addlegendentry{Mixed}

\addplot [color=color3,semithick]
  table[row sep=crcr]{%
0	96.3733086333668\\
50	96.3733086333668\\
};
\addlegendentry{Pure sub-6~GHz}

\addplot [color=color5,semithick, mark=o, mark options={solid, color5}]
  table[row sep=crcr]{%
0 100\\
2.5	99.9375000000002\\
5	99.7500000000012\\
7.5	99.4375000000002\\
10	99.0000000000013\\
12.5	98.4375000000011\\
15	97.7500000000031\\
17.5	96.9375000000034\\
20	96.0000000000049\\
22.5	94.9375000000048\\
25	93.7500000000076\\
27.5	92.4375000000078\\
30	91.000000000009\\
32.5	89.4375000000098\\
35	87.7500000000136\\
37.5	85.9375000000165\\
40	84.0000000000188\\
42.5	81.9375000000197\\
45	79.7500000000215\\
47.5	77.4375000000254\\
50	75.0000000000276\\
};
\addlegendentry{Pure mmWave}
\end{axis}

\end{tikzpicture}%
        \caption{Latency-constrained reliability as a function of the outage probability of the mmWave links. The capacity of the sub-6 Link is fixed to $36\text{Mbit/s}$.}
        \label{fig:pout}
\end{figure}

While relatively common in the literature, mixed systems using a sub-6 link as a backup for a high-capacity mmWave link actually present a problem: since outages in the mmWave link result in the delivery of no packets, the sub-6 alternative needs to be able to deliver the whole block on its own within the deadline to be an effective backup. Fig.~\ref{fig:pout} shows the latency-constrained reliability of the systems in the three considered cases as a function of the outage probability $p_{\text{out}}$ for the mmWave links. Naturally, having 2 parallel mmWave links works best if the outage probability is low, as the frame can be replicated fully and sent over both. However, if $p_{\text{out}}\geq0.2$, the two mmWave links will often be blocked at the same time, and the pure sub-6 case, in which both links deliver at least part of the frame and packet-level coding is fully exploited, actually provides a better performance. Interestingly, mixing a mmWave link and a sub-6 link is not the best choice for any value of the outage probability, as another mmWave link would be better if $p_{\text{out}}$ is low, and the pure sub-6 scheme can use packet-level coding more efficiently if $p_{\text{out}}$ is large.

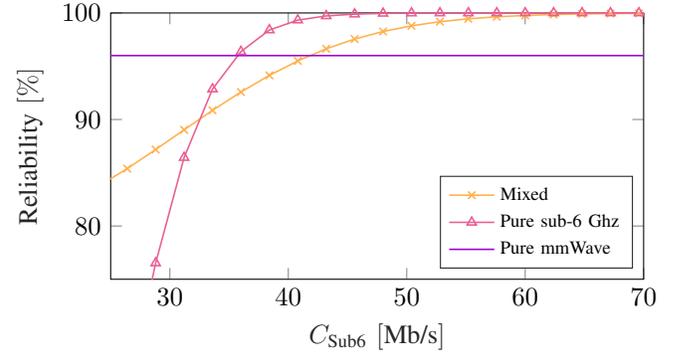
\begin{figure}[t]
  \centering
%
%
\definecolor{mycolor1}{rgb}{0.00000,0.44700,0.74100}%
\definecolor{mycolor2}{rgb}{0.85000,0.32500,0.09800}%
\definecolor{mycolor3}{rgb}{0.92900,0.69400,0.12500}%
\begin{tikzpicture}

\begin{axis}[%
width=\fwidth,
height=\fheight,
scale only axis,
xmin=25,
xmax=70,
xlabel style={font=\color{white!15!black}},
xlabel={${C}_{\text{Sub6}}\ [\text{Mb/s}]$},
ymin=75,
ymax=100,
ylabel style={font=\color{white!15!black}},
ylabel={Reliability $[\%]$},
axis background/.style={fill=white},
legend style={legend cell align=left, align=left, draw=white!15!black, at={(0.98,0.04)},anchor=south east,font={\scriptsize}}
]
\addplot [color=color1, semithick, mark=x, mark options={solid, color1}]
  table[row sep=crcr]{%
24	83.7641670984264\\
26.4	85.3878953842607\\
28.8	87.1799851486976\\
31.2	89.0306455429364\\
33.6	90.8597623641404\\
36	92.5801628706075\\
38.4	94.1399769297375\\
40.8	95.4998708324246\\
43.2	96.6365005231423\\
45.6	97.5513696688328\\
48	98.2614597453249\\
50.4	98.7936203431997\\
52.8	99.1801152064939\\
55.2	99.4530178930854\\
57.6	99.6409819492503\\
60	99.7676753240259\\
62.4	99.8514989444545\\
64.8	99.9060864356907\\
67.2	99.9411582960446\\
69.6	99.9634335904681\\
72	99.9774419661832\\
};
\addlegendentry{Mixed}

\addplot [color=color3, semithick, mark=triangle, mark options={solid, color3}]
  table[row sep=crcr]{%
24	51.4443555642179\\
26.4	64.1598240492498\\
28.8	76.5243961169264\\
31.2	86.4233159084667\\
33.6	92.8540793954459\\
36	96.3733086333667\\
38.4	98.399751923817\\
40.8	99.3292598640163\\
43.2	99.7245033850643\\
45.6	99.8882755470644\\
48	99.9618535860849\\
50.4	99.9874704359998\\
52.8	99.9957329482639\\
55.2	99.9985336794697\\
57.6	99.9995994601765\\
60	99.9998914336332\\
62.4	99.9999683344971\\
64.8	99.9999898352008\\
67.2	99.9999974862058\\
69.6	99.999999408274\\
72	99.999999855593\\
};
\addlegendentry{Pure sub-6~Ghz}

\addplot [semithick, color=color5]
  table[row sep=crcr]{%
24	96.0000000000049\\
72	96.0000000000049\\
};
\addlegendentry{Pure mmWave}

\end{axis}

\end{tikzpicture}%
        \caption{Latency-constrained reliability as a function of the average capacity of the sub-6~GHz links. The mmWave link outage probability is $p_{\text{out}}=0.2$.}
        \label{fig:csub}
\end{figure}

\begin{figure}[t]
     \centering
        \begin{tikzpicture}[scale=0.8]
   \node[minimum size=6mm, font=\small,text=black] at (4,-0.5) {Link 1  queue};
  \node[minimum size=6mm, rotate=90, font=\small,text=black] at (0.5,-4) {Link 2 queue};
 
 \foreach \x in {0,...,4} {
  \node[fill=white, minimum size=6mm, font=\scriptsize,text=black] at (\x+2,-1) {\x};
  \node[fill=white, minimum size=6mm, font=\scriptsize,text=black] at (1,-\x-2) {\x};
 }
\foreach \y [count=\n] in {{0.100,0.100,0.200,0.100,0.200},
{0.100,0.200,0.100,0.200,0.100},
{0.200,0.100,0.200,0.100,0.100},
{0.100,0.200,0.100,0.100,0.100},
{0.200,0.100,0.100,0.100,0.00}}
 {
  \foreach \x [count=\m] in \y {
   \pgfmathsetmacro\result{\x * 90}
   \ifnum\pdfstrcmp{\x}{-1.000}=0
    \node[fill=white!80!red, minimum size=8mm, text=black] at (\m+1,-\n-1) { \centering \tiny drop };
   \else
    \node[fill=blue!\result!white, minimum size=8mm, text=black] at (\m+1,-\n-1) { \centering \tiny};
   \fi
  }
 }
 
\foreach \y [count=\n] in {{0.6\\0.5,0.5\\0.6,0.5\\0.7,0.4\\0.7,0.4\\0.8},
{0.6\\0.5,0.6\\0.6,0.5\\0.6,0.5\\0.7,0.4\\0.7},
{0.7\\0.5,0.6\\0.5,0.6\\0.6,0.5\\0.6,0.4\\0.7},
{0.7\\0.4,0.7\\0.5,0.6\\0.5,0.6\\0.5,0.5\\0.6},
{0.8\\0.4,0.7\\0.4,0.7\\0.4,0.6\\0.5,0.5\\0.5}}
 {
  \foreach \x [count=\m] in \y {
    \node[fill=none, minimum size=10mm, text=black] at (\m+1,-\n-1) { \centering \scriptsize \makecell[c]{\x}};
  }
 }
\end{tikzpicture}
         \caption{Strategies for the pure sub-6~GHz system.}
         \label{fig:str_sub6}
\end{figure}

We can see the same if we fix $p_{\text{out}}=0.2$ and vary the capacity of the sub-6 links, as we show in Fig.~\ref{fig:csub}: the pure mmWave case has a reliability slightly higher than 95\%, but two sub-6 channels can do better as soon as their capacity greater than about 35~Mb/s. The mixed case is only better than the pure sub-6 if the capacity is below 30~Mb/s, with the mmWave link taking most of the responsibility for the reliable delivery.

We can also consider the different strategies for the three scenarios. In the pure mmWave system, the optimal strategy is always to send a complete copy of the frame if the link is not in outage and to avoid sending anything otherwise, to avoid filling the queue. However, even creating a queue is not a problem, as the mmWave links have a significantly higher capacity than what is needed. Fig. \ref{fig:str_sub6} and Fig. \ref{fig:str_mix} show the strategies for the other two cases, with $p_{\text{out}}=0.2$ and $C_{\text{Sub6}}=36$~Mb/s. The figures show the fraction of the data sent over each link as a function of the number of packets in the two transmission queues: if we use redundant coding, the sum is larger than 1. The cells are also colored based on the amount of redundancy, with darker cells corresponding to a more robust encoding. 

As expected, the strategy for the pure sub-6 case, shown in Fig.~\ref{fig:str_sub6}, is symmetrical: since the two links are identical, the optimal scheduling is mirrored if we reverse the states of the queues. The granularity of the optimization was 0.1, i.e., we considered blocks of 40~kb for the encoding, which explains the apparent asymmetry when both queues are empty. In general, the added redundancy is between 10\% and 20\%, and we can notice that, in highly unbalanced cases, the link with a shorter queue tends to be assigned more packets, allowing the more congested link to transmit fewer packets and flush the queue. We only showed results for queues of up to 4 packets, as other cases occur with extremely low probability: the optimal strategy tries to maintain short queues, sacrificing some reliability to avoid snowball effects in the near future. This is clear from the behavior when 4 packets are in each queue, in which case no redundancy is sent: in more extreme unlucky cases, the frame might even be dropped entirely to allow the links to reduce their queues and improve the situation for future frames.

\begin{figure}[t]
         \centering
        \begin{tikzpicture}[scale=0.8]
  \node[minimum size=6mm, font=\small,text=black] at (4,-0.5) {mmWave queue};
  \node[minimum size=6mm, rotate=90, font=\small,text=black] at (0.5,-4) {Sub-6~GHz queue};

 \foreach \x in {0,...,4} {
  \node[fill=white, minimum size=6mm, font=\scriptsize,text=black] at (\x+2,-1) {\x};
  \node[fill=white, minimum size=6mm, font=\scriptsize,text=black] at (1,-\x-2) {\x};
 }
\foreach \y [count=\n] in {{0.200,0.200,0.300,0.300,0.400},
{0.100,0.200,0.200,0.200,0.200},
{0.100,0.100,0.100,0.100,0.100},
{0.000,0.000,0.000,0.000,0.000},
{0.000,0.000,0.000,0.000,0.000}}
 {
  \foreach \x [count=\m] in \y {
   \pgfmathsetmacro\result{\x * 60}
   \ifnum\pdfstrcmp{\x}{-1}=0
    \node[fill=white!80!red, minimum size=8mm, text=black] at (\m+1,-\n-1) { \centering \tiny drop };
   \else
    \node[fill=blue!\result!white, minimum size=8mm, text=black] at (\m+1,-\n-1) { \centering \tiny};
   \fi
  }
 }

  \foreach \y [count=\n] in {{1\\0.2,1\\0.2,1\\0.3,1\\0.3,1\\0.4},
{1\\0.1,1\\0.2,1\\0.2,1\\0.2,1\\0.2},
{1\\0.1,1\\0.1,1\\0.1,1\\0.1,1\\0.1},
{1\\0,1\\0,1\\0,1\\0,1\\0},
{1\\0,1\\0,1\\0,1\\0,1\\0}}
 {
  \foreach \x [count=\m] in \y {
    \node[fill=none, minimum size=10mm, text=black] at (\m+1,-\n-1) { \centering \scriptsize \makecell[c]{\x}};
  }
 }
\end{tikzpicture}
 
         \caption{Strategies for the mixed system (not in outage).}
         \label{fig:str_mix}
\end{figure}

We can also look at the optimal policy in the mixed case, shown in Fig.~\ref{fig:str_mix} when the mmWave link is active: in all cases, all packets are transmitted over this link, while the sub-6 link is only used to transmit additional redundancy. In general, the probability of the system being in one of the states on the right (i.e., of the mmWave link having a significant queue) is extremely low, and these states are only visited after outages. In case of outage, the situation is reversed, as the sub-6 link is used to transmit all packets, while the mmWave link is left unused to avoid building up queues. As we mentioned above, this would be different with an entirely end-to-end capacity estimation: in that case, the transmitter would need to send some probe packets to avoid wasting capacity due to starvation.

In general, the basic principle of the optimal policy is always to use the strongest links first, aiming at avoiding building up queues, but finding the correct amount of redundancy and the correct threshold for shifting the load between links is tricky even in this simple scenario. However, the optimal strategy also manages to strengthen the reliability of links, using the independence between channels to compensate for delays and losses even when the application has no control over the underlying network.

\section{Conclusions and Future Work}\label{sec:conc}

In this paper, we have presented multipath wireless communications with packet-level coding over multiple links as a possible solution to provide reliable low-latency service to teleoperation and other novel high-throughput industrial applications. We described in detail the main issues and challenges of such a setup, with considerations on different levels of control over the underlying network and for different wireless technologies. We also presented an \gls{mdp} model to derive the optimal coding and scheduling policy in a generic network, and give a simple but realistic example. 

As \gls{mdp} models are an active topic of research with quick improvements on new \gls{rl} solutions, the most interesting research direction for communications is the design of the model itself: finding a compact but expressive representation for the state of a link, which might be over different technologies and with different levels of application control over the scheduler and queues, is a significant challenge. The investigation of partially correlated links, such as mmWave with geometric blockers, is also an extremely interesting extension of our idea, which changes the stakes in allocating redundant information.

\bibliographystyle{IEEEtran}
\bibliography{./bibliography.bib}

\begin{thebibliography}{10}
\providecommand{\url}[1]{#1}
\csname url@samestyle\endcsname
\providecommand{\newblock}{\relax}
\providecommand{\bibinfo}[2]{#2}
\providecommand{\BIBentrySTDinterwordspacing}{\spaceskip=0pt\relax}
\providecommand{\BIBentryALTinterwordstretchfactor}{4}
\providecommand{\BIBentryALTinterwordspacing}{\spaceskip=\fontdimen2\font plus
\BIBentryALTinterwordstretchfactor\fontdimen3\font minus
  \fontdimen4\font\relax}
\providecommand{\BIBforeignlanguage}[2]{{%
\expandafter\ifx\csname l@#1\endcsname\relax
\typeout{** WARNING: IEEEtran.bst: No hyphenation pattern has been}%
\typeout{** loaded for the language `#1'. Using the pattern for}%
\typeout{** the default language instead.}%
\else
\language=\csname l@#1\endcsname
\fi
#2}}
\providecommand{\BIBdecl}{\relax}
\BIBdecl

\bibitem{saad2019vision}
W.~Saad, M.~Bennis, and M.~Chen, ``A vision of {6G} wireless systems:
  Applications, trends, technologies, and open research problems,'' \emph{IEEE
  Network}, vol.~34, no.~3, pp. 134--142, 2019.

\bibitem{TS22104}
3GPP, ``Service requirements for cyber-physical control applications in
  vertical domains,'' Dec. 2021, {TS} 22.104.

\bibitem{hu2021vision}
F.~Hu, Y.~Deng, H.~Zhou, T.~H. Jung, C.-B. Chae, and A.~H. Aghvami, ``A vision
  of an {XR}-aided teleoperation system toward {5G/B5G},'' \emph{IEEE
  Communications Magazine}, vol.~59, no.~1, pp. 34--40, Feb.

\bibitem{gholipoor2020cloud}
N.~Gholipoor, S.~Parsaeefard, M.~R. Javan, N.~Mokari, H.~Saeedi, and
  H.~Pishro-Nik, ``Cloud-based queuing model for tactile internet in next
  generation of {RAN},'' in \emph{91st Vehicular Technology Conference
  (VTC2020-Spring)}.\hskip 1em plus 0.5em minus 0.4em\relax IEEE, May 2020.

\bibitem{ieee2021vrstandard}
``Standard for {Head-Mounted Display (HMD)-based Virtual Reality(VR)} sickness
  reduction technology,'' \emph{{IEEE Std. 3079-2020}}, Apr. 2021.

\bibitem{nielsen2017ultra}
J.~J. Nielsen, R.~Liu, and P.~Popovski, ``Ultra-reliable low latency
  communication using interface diversity,'' \emph{IEEE Transactions on
  Communications}, vol.~66, no.~3, pp. 1322--1334, Nov. 2017.

\bibitem{chiariotti2021hop}
F.~Chiariotti, A.~Zanella, S.~Kucera, K.~Fahmi, and H.~Claussen, ``The {HOP}
  protocol: Reliable latency-bounded end-to-end multipath communication,''
  \emph{IEEE/ACM Transactions on Networking}, vol.~29, no.~5, pp. 2281--2295,
  Jun. 2021.

\bibitem{howard1960dynamic}
R.~A. Howard, \emph{Dynamic programming and {Markov} processes}.\hskip 1em plus
  0.5em minus 0.4em\relax John Wiley, 1960.

\bibitem{zhang2019will}
M.~Zhang, M.~Polese, M.~Mezzavilla, J.~Zhu, S.~Rangan, S.~Panwar, and M.~Zorzi,
  ``Will {TCP} work in {mmWave 5G} cellular networks?'' \emph{IEEE
  Communications Magazine}, vol.~57, no.~1, pp. 65--71, Jan. 2019.

\bibitem{polese2019survey}
M.~Polese, F.~Chiariotti, E.~Bonetto, F.~Rigotto, A.~Zanella, and M.~Zorzi, ``A
  survey on recent advances in transport layer protocols,'' \emph{IEEE
  Communications Surveys \& Tutorials}, vol.~21, no.~4, pp. 3584--3608, Aug.
  2019.

\bibitem{shah2016redundant}
N.~B. {Shah}, K.~{Lee}, and K.~{Ramchandran}, ``When do redundant requests
  reduce latency?'' \emph{IEEE Transactions on Communications}, vol.~64, no.~2,
  pp. 715--722, Dec. 2015.

\bibitem{atxutegi2018use}
E.~Atxutegi, F.~Liberal, H.~K. Haile, K.-J. Grinnemo, A.~Brunstrom, and
  A.~Arvidsson, ``On the use of {TCP BBR} in cellular networks,'' \emph{IEEE
  Communications Magazine}, vol.~56, pp. 172--179, Mar. 2018.

\bibitem{xiang2021recent}
X.~Xiang, S.~Foo, and H.~Zang, ``Recent advances in {Deep Reinforcement
  Learning} applications for solving {Partially Observable Markov Decision
  Processes (POMDP)} problems part 2—applications in transportation,
  industries, communications and networking and more topics,'' \emph{MDPI
  Machine Learning and Knowledge Extraction}, vol.~3, no.~4, pp. 863--878, Jul.
  2021.

\bibitem{akyazi2018comparison}
P.~Akyazi and T.~Ebrahimi, ``Comparison of compression efficiency between
  {HEVC/H. 265, VP9 and AV1} based on subjective quality assessments,'' in
  \emph{10th International Conference on Quality of Multimedia Experience
  (QoMEX)}.\hskip 1em plus 0.5em minus 0.4em\relax IEEE, 2018.

\bibitem{liu2018mec}
Y.~Liu, J.~Liu, A.~Argyriou, and S.~Ci, ``{MEC}-assisted panoramic {VR} video
  streaming over millimeter wave mobile networks,'' \emph{IEEE Transactions on
  Multimedia}, vol.~21, no.~5, pp. 1302--1316, Oct. 2018.

\end{thebibliography}

\begin{IEEEbiography}[{\includegraphics[width=0.99in,clip,keepaspectratio]{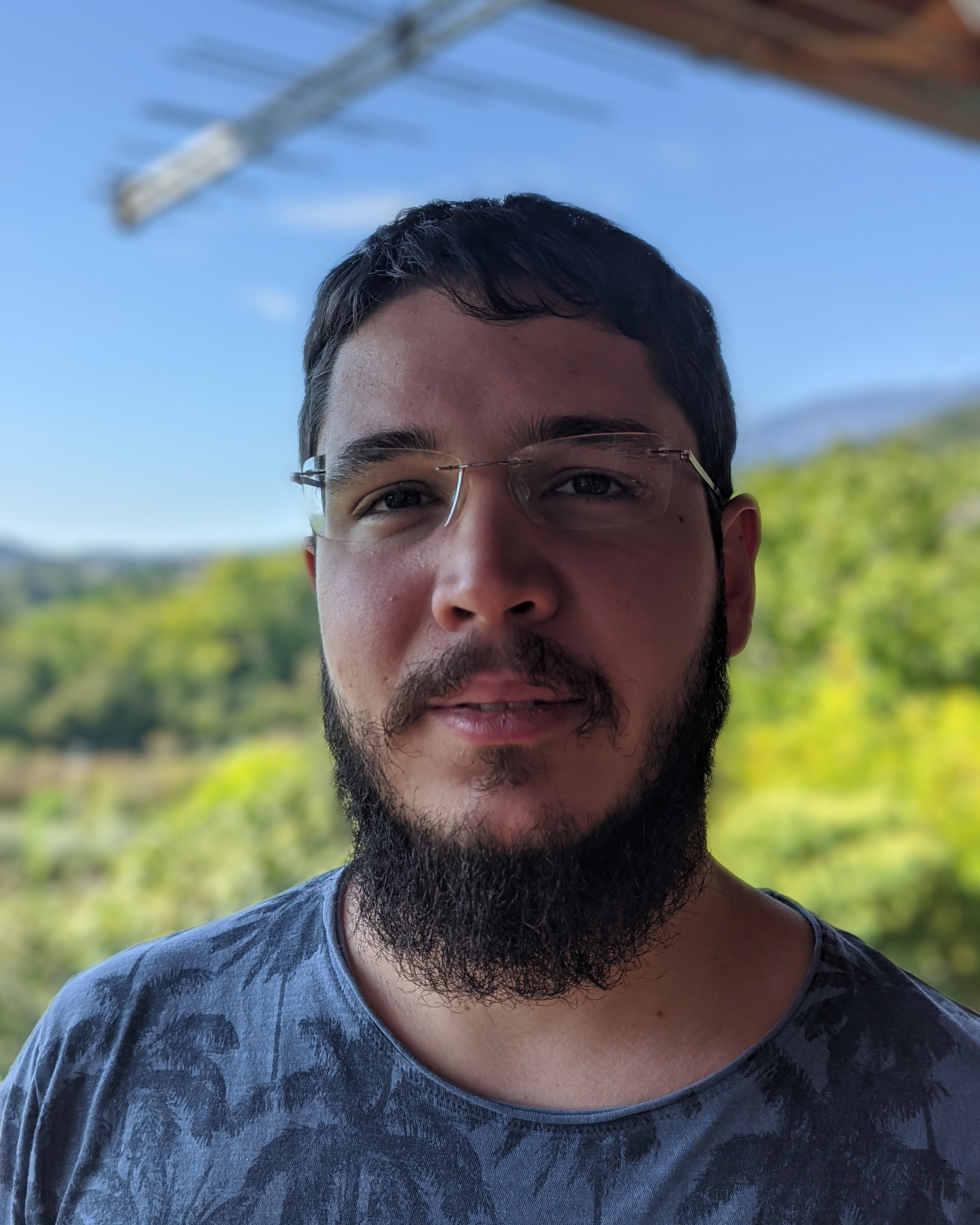}}]{Andrea Bedin} is an Early Stage Researcher within the MINTS project, currently working at Nokia Bell Labs in Espoo, Finland. He is also enrolled in the Ph.D. program in information engineering at the University of Padova. He obtained his master's degree in ICT for internet and multimedia engineering in 2020 and his bachelor's degree in information engineering in 2018 from the University of Padova. His research interests are in low latency ultra-reliable communications for industrial applications.
\end{IEEEbiography}%

\begin{IEEEbiography}[{\includegraphics[width=0.99in,clip,keepaspectratio]{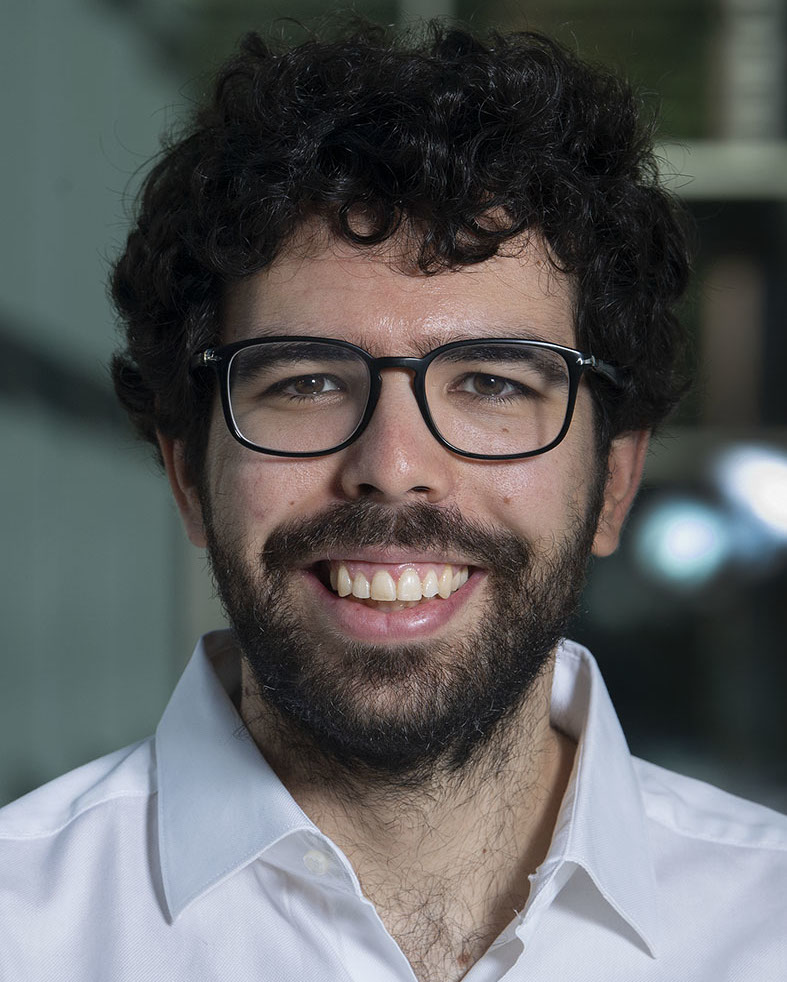}}]{Federico Chiariotti}  [S'15-M'19] is currently an Assistant Professor in the Department of Electronic Systems, Aalborg University, Denmark. He received his PhD in Information Engineering in 2019 from the University of Padova, Italy. He has published more than 50 peer-reviewed papers and received the Best Paper Award in 4 conferences. His research focuses on the latency-oriented design of networking protocols, the use of machine learning in network, and on goal-oriented communications.
\end{IEEEbiography}

\begin{IEEEbiography}[{\includegraphics[width=0.99in,clip,keepaspectratio]{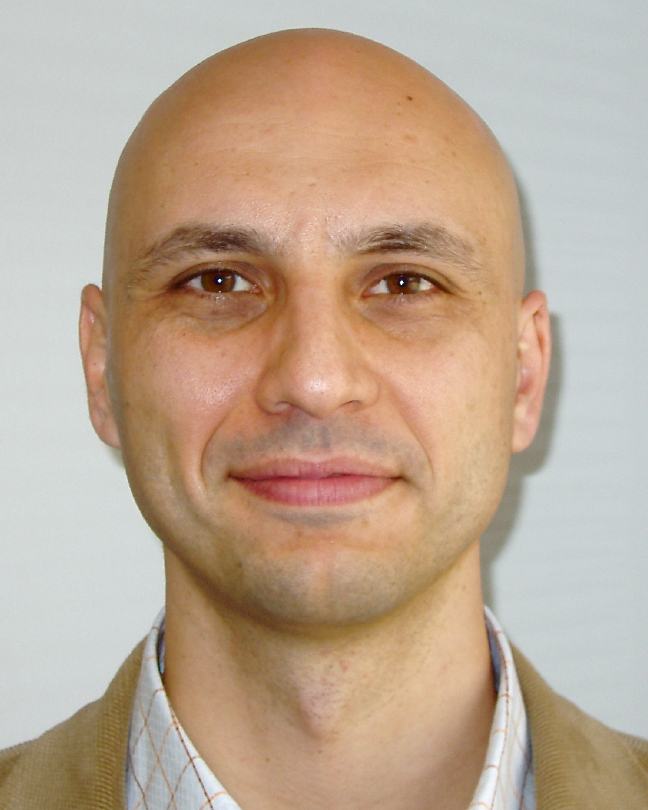}}]{Andrea Zanella} [S'98-M'01-SM'13] is a Full Professor at the Department of Information Engineering (DEI), University of Padova (Italy). He received the Laurea degree in Computer Engineering in 1998 and the PhD in 2001 from the same University. He is one of the coordinators of the SIGnals and NETworking (SIGNET) research lab. His long-established research activities are in the fields of protocol design, optimization, and performance evaluation of wired and wireless networks.
\end{IEEEbiography}%

\end{document}